\documentclass[showkeys,12pt, preprint,preprintnumbers,nofootinbib,
groupedaddress,superscriptaddress,amsmath,amssymb]{revtex4} 
\usepackage{graphicx}
\usepackage{dcolumn}
\usepackage{bm}
\usepackage{amssymb}
\usepackage{amsmath}
\usepackage{epsfig}
\usepackage{color}
\usepackage{slashed}
\usepackage{hhline}

\def\be{\begin{equation}}
\def\ee{\end{equation}}
\newcommand{\bea}{\begin{eqnarray}}
\newcommand{\eea}{\end{eqnarray}}

\numberwithin{equation}{section}

\begin{document}

{\begin{flushright}{KIAS-P17012}\end{flushright}}

\title{
Discriminating leptonic Yukawa interactions with doubly charged scalar
at the ILC}

\author{Takaaki Nomura}
\email{nomura@kias.re.kr}
\affiliation{School of Physics, KIAS, Seoul 02455, Korea}

\author{Hiroshi Okada}
\email{macokada3hiroshi@cts.nthu.edu.tw}
\affiliation{Physics Division, National Center for Theoretical Sciences,
Hsinchu, Taiwan 300}

\author{Hiroshi Yokoya}
\email{hyokoya@kias.re.kr}
\affiliation{Quantum Universe Center, KIAS, Seoul 02455, Korea}

\date{\today}

\begin{abstract}
 We explore discrimination of two types of leptonic Yukawa interactions
 associated with Higgs triplet, $\bar L^c_L \Delta L_L$, and with
 $SU(2)$ singlet doubly charged scalar, $\bar e_R^c k^{++} e_R$.
 These interactions can be distinguished by measuring the effects of
 doubly charged scalar boson exchange in the $e^+ e^- \to \ell^+ \ell^-$
 processes at polarized electron-positron colliders.
 We study a forward-backward asymmetry of scattering angular
 distribution to estimate the sensitivity for these effects at the
 ILC.
 In addition, we investigate prospects of upper bounds on the Yukawa
 couplings by combining the constraints of lepton flavor violation
 processes and the $e^+ e^- \to \ell^+ \ell^-$ processes at the LEP and
 the ILC.
\end{abstract}

\maketitle

\newpage

\section{Introduction}

Doubly charged scalar bosons sometimes play an important role in
building models which explain neutrino masses and mixing.
One of the representative scenarios is called type-II seesaw
mechanism~\cite{Magg:1980ut, Konetschny:1977bn,Cheng:1980qt} in which an isospin
triplet scalar field $\Delta$ is introduced. 
Since $\Delta$ has $U(1)_Y$ hypercharge $Y=1$~\footnote{Here we apply the convention for hypercharge where electric charge is given by $Q = Y + T_3$ and the diagonal generator of $SU(2)$ gauge symmetry $T_3$ has eigenvalues $\{1,0,-1\}$ for a triplet. }, it contains an
electrically neutral~($\delta^0$), singly-charged~($\delta^\pm$), and
doubly charged~($\delta^{\pm\pm}$) bosons, and directly couples to the
isospin-doublet leptons at tree level via $h_{ij} \bar L_{L_i}^c
\Delta L_{L_j}$. 
The neutrino mass matrix is induced after $\Delta$ develops a
vacuum expectation value~(VEV) which violates the lepton number by two.
To achieve the smallness of neutrino masses, the VEV of
$\Delta$ or/and the Yukawa couplings $h_{ij}$ have to be
small.\footnote{%
Notice here that the upper bound on the triplet VEV is about a few GeV,
which is derived by the fact that the deviation of the electroweak rho
parameter from unity is severely constrained by experimental
measurements.}
Interactions of $\bar\ell_L^c \ell_L\delta^{++}$ and $\bar\nu_L^c
\ell_L\delta^{+}$ provide promising signals of doubly~(singly) charged
bosons at collider experiments.
Notice here that these interactions are associated with leptons with
left-handed chirality.

Another type of interaction is often employed in radiative seesaw models,
such as the Zee-Babu model~\cite{2-lp-zB}, in which doubly (singly)
charged scalar bosons, $k^{\pm\pm}~(h^\pm)$, are introduced as an
isospin singlet; {the topology of neutrino mass generating diagram was shown in Ref.~\cite{Cheng:1980qt} for the first time}. 
Then $k^{\pm \pm}$ directly couples to the isospin-singlet leptons at
tree level via $h'_{ij} \bar e_{R_i}^c k^{++} e_{R_j}$~\footnote{This type of interaction also appears in left-right symmetric models~\cite{Mohapatra:1974gc, Pati:1974yy, Senjanovic:1975rk, Senjanovic:1978ev, Mohapatra:1979ia, Mohapatra:1980yp} associated with triplet scalar which couples to right-handed fermions. }, which is
also a source of lepton number violation by two unit. 
The neutrino mass matrix is generated at two-loop level via
several terms related to the doubly~(singly) charged bosons after the
electroweak symmetry breaking by the standard-model~(SM) Higgs boson
where we can naturally consider models with ${\cal O}(1)$ Yukawa
coupling and/or new particles whose masses are ${\cal O}(0.1-1)$~TeV
inside the loops.
The interaction is associated with leptons
with right-handed chirality in contrast to the type-II seesaw scenario. 

These two types of Yukawa interaction would also appear in
many other neutrino mass models as a source of lepton number violation.
Thus it is important to analyze the interactions of doubly charged
scalar bosons and charged leptons to discriminate models of neutrino
mass generation~\cite{Sugiyama:2012yw}.
{ Searches for the doubly charged scalar bosons have been performed in
the past and current collider experiments, and the most severe
constraint is obtained by the same-sign di-lepton resonance searches at
the LHC~\cite{ Aad:2012cg,ATLAS:2014kca,ATLAS:2016pbt,Chatrchyan:2012ya, Biondini:2014dfa, Babu:2016rcr, ATLAS:2017iqw}.
In addition, multi lepton signals are discussed including doubly charged Higgs production associated with singly charged Higgs~\cite{Akeroyd:2010ip, CMS:2017pet}.
 Assuming a large branching ratio of a certain di-lepton channel, 
the most stringent bound on the mass of doubly charged scalar bosons varies from
770~GeV to 870~GeV depending on the models and the decay branching ratio, which is obtained from the latest LHC data~\cite{ATLAS:2017iqw}.}

The two bosons have also different interaction with the electroweak gauge
bosons.
Namely the doubly charged scalar boson of triplet models can couple to
a pair of charged gauge bosons, $W^\pm$'s, at tree level.
The decay branching ratio to the same-sign $WW$ can be large if the
triplet VEV is as large as ${\cal O}(0.1-10)$~MeV depending on the mass
itself.
Thus the search for such events provides an unique signature of the
triplet model~\cite{%
Kanemura:2013vxa,Kanemura:2014goa,Kanemura:2014ipa}. 
Production of doubly charged scalar bosons via the $WW$-fusion process
is also expected in the triplet-like models, such as the Georgi-Machacek
model~\cite{Georgi:1985nv}, which realize a large triplet VEV to make
the production cross section sizable~\cite{%
Chiang:2012dk,Dutta:2014dba,Khachatryan:2014sta}.

In this paper, we study the effects of the doubly charged
scalar boson exchange in $e^+e^-\to\ell^+\ell^-$
processes~\cite{Swartz:1989qz,Eichten:1983hw} at future lepton colliders.
We consider the initial-state polarization to discriminate the two types
of the interactions, and study the bounds of their masses and couplings
probed at the future international linear collider~(ILC)
experiment~\cite{Baer:2013cma}.
Furthermore, we discuss prospects of constraining model parameter space
by combining the current and future constraints of lepton flavor
violating~(LFV) processes as well as the constraints at the LEP
experiment~\cite{Schael:2013ita}.

The paper is organized as follows.
In Sec.~II, we discuss constraints from LFV processes. 
In Sec.~III, we consider the effects of doubly charged scalar bosons to
the $e^+e^-\to\ell^+\ell^-$ processes at lepton colliders, and study the
constraints on the mass and couplings of these bosons by using the
forward-backward asymmetry.
In Sec.~IV we summarize the constraints and prospects of future bounds
on the relevant effective couplings with numerical analysis. 
Sec.~V is devoted for conclusions and discussions.

\section{ LFV processes}
{ In this section, constrains from LFV processes are reviewed where we summarize bounds of LFV processes on the two representative
terms, $\bar L_{L}^c(i\sigma_2) \Delta L_{L}$ in Higgs triplet models, and $\bar e_R^c k^{++} e_R$ in Zee-Babu type models. }

\subsection{ LFV processes in a Higgs triplet model}
We discuss the LFV processes in a Higgs triplet model induced by
following relevant Lagrangian\footnote{%
A notation in this paper is different from that in some papers such as
Ref.~\cite{Okada:2015gia}  where $\sqrt2$ is replaced by
$1/\sqrt2$ in Eq.~(II.1).
\label{fn:convention}}:
\begin{align}
&-{\cal L}= h_{ij} \bar L_{L_i}^c(i\sigma_2) \Delta L_{L_j} {+ {\rm h.c.}} \supset
 -h_{ij}\left[\sqrt2 \bar \nu^c_{L_i} \ell_{L_j}\delta^{+} + \bar
 \ell^c_{L_i} \ell_{L_j}\delta^{++} \right]+{\rm h.c.},
 \label{eq:type-II}\\ 
&L_L =\left[
\begin{array}{c}
\nu_L \\
 \ell_L
\end{array}\right],
\quad \Delta =\left[
\begin{array}{cc}
\frac{\delta^+}{\sqrt2} & \delta^{++}\\
\delta^{0} & -\frac{\delta^+}{\sqrt2}
\end{array}\right],
\end{align}
where $h_{ij}$ is a symmetric Yukawa coupling matrix, $h_{ij}=h_{ji}$,
whose indices run over ($e$, $\mu$, $\tau$),
and $\sigma_2$ is the second Pauli matrix. Thus the doubly charged Higgs interactions are left-handed type in this case.

\begin{figure}[t]
\begin{center}
\includegraphics[width=50mm]{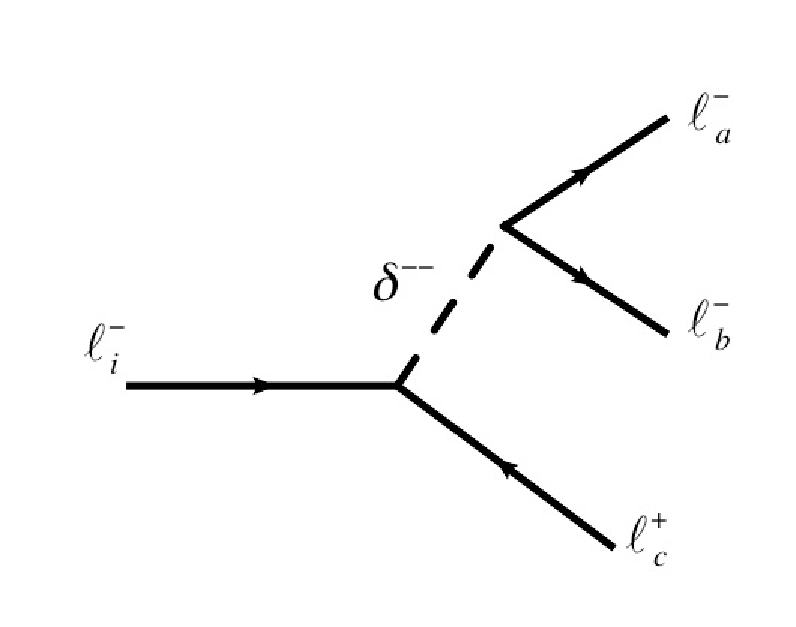} 
\includegraphics[width=80mm]{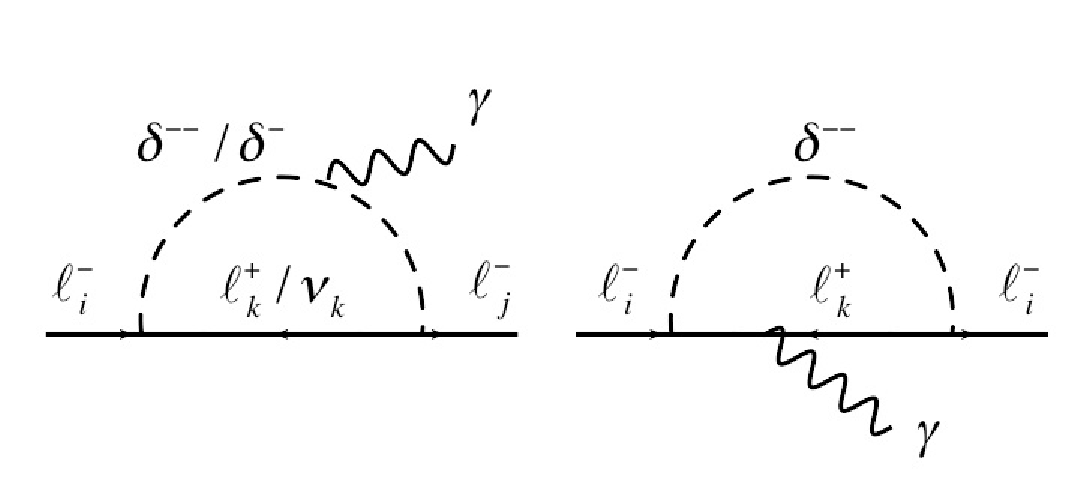} 
 \caption{Left: A diagram for $\ell^-_i \to \ell^-_a \ell^-_b \ell^+_c$ process via doubly charged scalar exchange. Right: One-loop diagrams for $\ell_i^- \to \ell_j^- \gamma$ process. }
 \label{fig:LFV}
\end{center}\end{figure}
These Yukawa interactions induce LFV processes $\ell_i^- \to \ell_j^-
\ell_k^+ \ell_l^-$ and $\ell_i \to \ell_j \gamma$ at tree and one-loop
levels, respectively, through virtual $\delta^{\pm \pm}$ and $\delta^\pm$
mediations {as shown in Fig.~\ref{fig:LFV}}. 
Then one obtains the constraints on the combination of $h_{ij}$ and
masses of $\delta^\pm$ and $\delta^{\pm\pm}$ by comparing the measured
branching ratios (BRs) of these LFV processes and those of model
predictions. 
Table~\ref{tab:LFVconst1} summarizes the {upper limits} of the BRs
and the corresponding constraints for the three-body decay
processes~\cite{Akeroyd:2009nu}, and Table~\ref{tab:LFVconst2}
summarizes the constraints from the upper limit of BR for $\mu\to
e\gamma$ given by the recent MEG experiment~\cite{TheMEG:2016wtm}, and
from the upper limits of BRs for $\tau \to \mu\gamma$, $e\gamma$ given
by Ref.~\cite{Adam:2013mnn} {applying results in Ref.~\cite{Okada:2015gia}}~\footnote{The constraints are obtained from the analysis in the reference where we take into account the difference of notations mentioned in footnote~\ref{fn:convention}.}.

\begin{table}[t]
\begin{tabular}{c|c} \hline
{Upper limits} of BRs & Constraints on $|h_{ij}^* h_{kl}|$ \\ \hline
~~~${\rm BR}(\mu^-\to e^+e^-e^-)\lesssim 1.0\times 10^{-12}$~~~ & ~~~$|h_{e\mu} h_{ee}^*|\lesssim 2.3\times 10^{-5}\times \left(\frac{{m_{\delta^{\pm\pm}}^{}}}{{\rm TeV}}\right)^2$~~~ \\
~~~${\rm BR}(\tau^-\to e^+e^-e^-) \lesssim2.7\times 10^{-8}$~~~ & ~~~$|h_{e\tau} h_{ee}^*|\lesssim 0.009\times \left(\frac{{m_{\delta^{\pm\pm}}^{}}}{{\rm TeV}}\right)^2$~~~~ \\
~~~${\rm BR}(\tau^-\to e^+e^-\mu^-) \lesssim1.8\times 10^{-8}$~~~ & ~~~$ |h_{e\tau} h_{e\mu}^*|\lesssim 0.005\times \left(\frac{{m_{\delta^{\pm\pm}}^{}}}{{\rm TeV}}\right)^2$~~~ \\
~~~${\rm BR}(\tau^-\to e^+\mu^-\mu^-) \lesssim1.7\times 10^{-8}$~~~ & ~~~$|h_{e\tau} h_{\mu\mu*}|\lesssim 0.007\times \left(\frac{{m_{\delta^{\pm\pm}}^{}}}{{\rm TeV}}\right)^2$~~~ \\
~~~${\rm BR}(\tau^-\to \mu^+e^-e^-) \lesssim1.5\times 10^{-8}$~~~ & ~~~$|h_{\mu\tau} h_{ee}^*|\lesssim 0.007\times \left(\frac{{m_{\delta^{\pm\pm}}^{}}}{{\rm TeV}}\right)^2$~~~~ \\
~~~${\rm BR}(\tau^-\to \mu^+e^-\mu^-) \lesssim2.7\times 10^{-8}$~~~ & ~~~$ |h_{\mu\tau} h_{e\mu}^*|\lesssim 0.007\times \left(\frac{{m_{\delta^{\pm\pm}}^{}}}{{\rm TeV}}\right)^2$~~~ \\
~~~${\rm BR}(\tau^-\to \mu^+\mu^-\mu^-) \lesssim2.1\times 10^{-8}$~~~ & ~~~$ |h_{\mu\tau} h_{\mu\mu}^*|\lesssim 0.008\times \left(\frac{{m_{\delta^{\pm\pm}}^{}}}{{\rm TeV}}\right)^2$~~~ \\ \hline
\end{tabular}
\caption{A summary of constraints on the combination of couplings
 $h_{ij}$ from LFV three-body decay branching ratios (BRs).} 
\label{tab:LFVconst1}
\end{table}
\begin{table}[t]
\begin{tabular}{c|c} \hline
{Upper limits} of BRs & Constraints  \\ \hline
~~~${\rm BR}(\mu\to e\gamma) \lesssim4.2\times 10^{-13}$~~~ & ~~~$\left|\sum_{i=e,\mu,\tau} h_{ie}^* h_{i\mu} \right|^2 \lesssim 4.72\times 10^{-6} \times \frac{(m_{\delta^{\pm}} m_{\delta^{\pm\pm}})^4}{(2m_{\delta^{\pm}}^2+m_{\delta^{\pm\pm}}^2)^2} \frac{1}{({\rm TeV})^4}$~~~  \\
~~~${\rm BR}(\tau\to e\gamma) \lesssim3.3\times 10^{-8}$~~~ & ~~~$ \left|\sum_{i=e,\mu,\tau} h_{ie}^* h_{i\tau} \right|^2  \lesssim 2.08\times 
 \frac{(m_{\delta^{\pm}} m_{\delta^{\pm\pm}})^4}{(2m_{\delta^{\pm}}^2+m_{\delta^{\pm\pm}}^2)^2} \frac{1}{({\rm TeV})^4}$~~~~ \\ 
~~~${\rm BR}(\tau\to \mu \gamma) \lesssim4.4\times 10^{-8}$~~~ & ~~~$\left|\sum_{i=e,\mu,\tau} h_{i\mu}^* h_{i\tau} \right|^2 \lesssim 2.85\times 
 \frac{(m_{\delta^{\pm}} m_{\delta^{\pm\pm}})^4}{(2m_{\delta^{\pm}}^2+m_{\delta^{\pm\pm}}^2)^2} \frac{1}{({\rm TeV})^4}$~~~~ \\ \hline 
\end{tabular}
\caption{A summary of constraints on the combination of couplings
 $h_{ij}$ from $\ell_i \to \ell_j \gamma$ decay branching ratios (BRs).} 
\label{tab:LFVconst2}
\end{table}

The Yukawa interactions also induce $\mu-e$ conversion via the same
one-loop diagrams which induce the $\mu \to e \gamma$ process. 
The $\mu-e$ conversion rate $R$ is approximately given by the following
form~\cite{Alonso:2012ji}: 
\begin{align}
R&\approx
\frac{4\alpha_{\rm em}^5Z_{\rm eff}^4F^2 m_\mu^5}{2^{10} 3^4\pi^4
 m_{\delta^{++}}^4\Gamma_{\rm capt}} 
\left|
\sum_{i=e,\mu,\tau} h_{ie}^* h_{i\mu} \left(
12+\frac{6m_{\delta^{++}}^2}{m_{\delta^{+}}^2}
+\frac{5+\epsilon_i(-22+5\epsilon_i)}{(-1+\epsilon_i)^3}
+\frac{6(-1+3\epsilon_i)\ln[\epsilon_i]}{(-1+\epsilon_i)^4}
\right)
\right|^2
 \label{eq:typeII_3},
\end{align}
where $\epsilon_i=m^2_i/m^2_{\delta^{++}}$, and nucleus-dependent
parameters $Z_{\rm eff}$, $F$, $\Gamma_{\rm capt}$ as well as the bounds
on $R$ are listed in Table~\ref{tab:mue-conv}.
Then the resultant upper bounds for the Yukawa coupling are estimated
once boson masses are fixed.
Moreover future experimental bounds for $N=({\rm Ti}, {\rm Al})$ can be
estimated~\cite{Barlow:2011zza, Hungerford:2009zz}. 

\begin{table}[t]
\begin{tabular}{c|c|c|c|c} \hline
Nucleus $^A_Z N$ & $Z_{\rm eff}$ & $F$ & $ \Gamma_{\rm capt}~[10^6{\rm sec}^{-1}]$ & Experimental bounds (Future bound) \\ \hline
$^{27}_{13}$Al & $11.5$ & $0.64$ & $0.7054$  & ($R_{\rm Al}\lesssim10^{-16}$)~\cite{Hungerford:2009zz} \\
$^{48}_{22}$Ti & $17.6$ & $0.54$ &$2.59$  & 
$R_{\rm Ti}\lesssim 4.3\times 10^{-12}$~\cite{Dohmen:1993mp}\ 
($\lesssim 10^{-18}$ \cite{Hungerford:2009zz}) \\
$^{197}_{79}$Au & $33.5$ & $0.16$ & $13.07$ & $R_{\rm Au}\lesssim7\times 10^{-13}$ ~\cite{Bertl:2006up}  \\ 
$^{208}_{82}$Pb & $34$ & $0.15$ & $13.45$   & $R_{\rm Pb}\lesssim4.6\times 10^{-11}$~\cite{Honecker:1996zf}  \\ \hline
\end{tabular}
\caption{
A summary of parameters for the $\mu-e$ conversion formula in various nuclei:
$Z_{\rm eff}$, $F\equiv|F_p(-m_\mu^2)|$, $\Gamma_{\rm capt}$~\cite{Alonso:2012ji},
 and the bounds on the capture rate $R$.}
\label{tab:mue-conv}
\end{table}

\subsection{ LFV processes with $\bar e^c_R k^{++}e_R$ interaction}
We proceed the same discussion as the Higgs triplet model in the case of  
\begin{align}
-{\cal L}= h'_{ij} \bar e^c_{R_i} k^{++}e_{R_j}+{\rm c.c.},
\label{eq:ZBtype}
\end{align}
where $h'_{ij}$ is also a symmetric Yukawa matrix; therefore,$h'_{ij}=h'_{ji}$. 
Thus the doubly charged Higgs interactions are right-handed type in this case.
Here notice that the constraint of the three-body decay $\ell_i\to
 \ell_j\ell_k\ell_\ell$ are the same as that for the Higgs triplet
 model by replacing $({h_{ij},\delta^{\pm\pm}}) \to
 ({h'_{ij},k^{\pm\pm}})$. 
On the other hand, the formulas for 
the constraint of $\ell_i\to\ell_j\gamma$ are different from those in
the Higgs triplet case due to the absence of the singly charged scalar boson.
Then the constraints on the combination of $h'_{ij}$ and the mass of
$k^{\pm\pm}$~\cite{Herrero-Garcia:2014hfa} are summarized in
Table~\ref{tab:LFVconst3}. 
\begin{table}[t]
\begin{tabular}{c|c} \hline
{Upper limits} of BRs & Constraints  \\ \hline
~~~${\rm BR}(\mu\to e\gamma) \lesssim4.2\times 10^{-13}$~~~ & ~~~$\left|\sum_{i=e,\mu,\tau} h_{ie}^{'*} h'_{i\mu} \right|^2 \lesssim 1.18\times 10^{-6}
\times \frac{m_{k^{\pm\pm}}^4}{({\rm TeV})^4}$~~~  \\
~~~${\rm BR}(\tau\to e\gamma) \lesssim3.3\times 10^{-8}$~~~ & ~~~$\left|\sum_{i=e,\mu,\tau} h_{ie}^{'*} h'_{i\tau} \right|^2  \lesssim 0.520\times 
 \frac{m_{k^{\pm\pm}}^4}{({\rm TeV})^4}$~~~~ \\ 
~~~${\rm BR}(\tau\to \mu \gamma) \lesssim4.4\times 10^{-8}$~~~ & ~~~$ \left|\sum_{i=e,\mu,\tau} h_{i\mu}^{'*} h'_{i\tau} \right|^2 \lesssim 0.713\times 
 \frac{m_{k^{\pm\pm}}^4}{({\rm TeV})^4}$~~~~ \\ \hline 
\end{tabular}
\caption{A summary of constraints on the combination of couplings $h'_{ij}$ from $\ell_i \to \ell_j \gamma$ decay branching ratios (BRs).}
\label{tab:LFVconst3}
\end{table}

The constraint of $\mu-e$ conversion is also given by the following
form: 
\begin{align}
R&\approx
\frac{4\alpha_{\rm em}^5Z_{\rm eff}^4F^2 m_\mu^5}{2^{10} 3^4\pi^4
 m_{k^{++}}^4\Gamma_{\rm capt}} 
\left|
\sum_{i=e,\mu,\tau} h_{ie}^{'*} h'_{i\mu} \left(
\frac{-7+\epsilon_i(-2+\epsilon_i)(-7+12\epsilon_i)}{(-1+\epsilon_i)^3}
+\frac{6(-1+3\epsilon_i)\ln[\epsilon_i]}{(-1+\epsilon_i)^4}
\right)
\right|^2
 \label{eq:ZBtype2},
\end{align}
where $Z_{\rm eff}$, $F$, $\Gamma_{\rm capt}$ and the bounds on $R$ are
given previously.

\section{Collider Physics}
Here we discuss effects of the two types of Yukawa interactions for
charged leptons and a doubly charged scalar boson in $e^+e^-$ collider
experiments.
The interactions of Eqs.~(\ref{eq:type-II}) and (\ref{eq:ZBtype})
contribute to the processes of $e^+ e^- \to \ell^+ \ell^-$ where the
doubly charged scalar bosons propagate in the
$u$-channel~\cite{Swartz:1989qz}. 
By taking the doubly charged scalar bosons as a heavy particle, we
obtain the following effective Lagrangians from the Higgs triplet and
Zee-Babu type interactions respectively: {
\begin{align}
\label{eq:effective1}
 {\cal L}_{\rm eff}^{\rm type-II}
&= \sum_{\ell = e,\mu,\tau} F_\ell
 \frac{|h_{e\ell}|^2}{m_{\delta^{\pm\pm}}^2} (\bar e\gamma^\mu
 P_Le)(\bar \ell\gamma_\mu P_L \ell)  \nonumber \\
 &\equiv \sum_{\ell = e,\mu,\tau} \frac{4\pi}{(1+\delta_{e\ell}) (\Lambda_L^\ell)^2} (\bar e\gamma^\mu
 P_Le)(\bar \ell\gamma_\mu P_L \ell), \\ 
 \label{eq:effective2}
 {\cal L}_{\rm eff}^{\rm Zee-Babu}
 &= \sum_{\ell = e,\mu,\tau} F_\ell \frac{|h'_{e\ell}|^2}{m_{k^{\pm\pm}}^2}
 (\bar e\gamma^\mu P_Re)(\bar \ell\gamma_\mu P_R \ell)  \nonumber \\
& \equiv  \sum_{\ell = e,\mu,\tau} \frac{4\pi}{(1+\delta_{e\ell}) (\Lambda_R^\ell)^2} (\bar e\gamma^\mu
 P_Re)(\bar \ell\gamma_\mu P_R \ell), 
\end{align}
where 
 $\delta_{e\ell}$ is the Kronecker delta, we used the Fierz transformation to get the relevant form of the
interactions, and $\{F_e, F_\mu, F_\tau\} = \{1/2, 2, 2\}$ is a factor coming from the symmetric structure of the Yukawa couplings: $h_{e \ell'} =
h_{\ell' e}(h'_{e \ell'}= h'_{\ell' e})$ for $\ell' = \mu, \tau$. 
Thus the scale parameters $\Lambda_{L,R}^\ell$ and original parameters in the models are related by:
\begin{align} 
\Lambda_{L}^\ell = \frac{\sqrt{2 (1+ \delta_{e\ell}) \pi}}{|h_{e\ell}|} m_{\delta^{\pm\pm}}, \quad
\Lambda_{R}^\ell = \frac{\sqrt{2 (1+\delta_{e \ell}) \pi}}{|h'_{e\ell}|} m_{k^{\pm\pm}}.
\label{eq:Lambda}
\end{align}}
Applying effective interactions Eq.~(\ref{eq:effective1}) and (\ref{eq:effective2}), we then show constraints by using
LEP results and prospects of exploring these interactions at the ILC.

Here, we emphasize merits of studying these scattering processes to
constrain the effective interactions, which are (1) a particular
coupling can be investigated in each process while combinations of
different couplings are constrained by LFV processes, (2) right-handed
and left-handed couplings can be clearly distinguished by using the
polarized electron and positron beams. 

\subsection{The constraints from LEP experiment}
{ The LEP constraints on the effective interactions are given in
Ref.~\cite{Schael:2013ita} in terms of $\Lambda_{L,R}^\ell$ which are obtained by fitting the scattering angular
distribution in $e^+e^-\to\ell^+\ell^-$ processes. 
We then rewrite the constraints for $\Lambda_{L,R}^\ell$ by the parameters in the original models using the relation Eq.~(\ref{eq:Lambda}).}
As a result the following constraints for each flavor $\ell$ are given in Higgs
triplet case: 
\begin{align}
 & |h_{ee}|\lesssim \frac{\sqrt{4\pi} m_{\delta^{\pm\pm}}}{8.7\ {\rm TeV}},\quad
 |h_{e\mu}|\lesssim \frac1{\sqrt2}\frac{\sqrt{4\pi}
 m_{\delta^{\pm\pm}}}{12.2\ {\rm TeV}},\quad 
 |h_{e\tau}|\lesssim \frac1{\sqrt2}\frac{\sqrt{4\pi}
 m_{\delta^{\pm\pm}}}{9.1\ {\rm TeV}},
 \end{align}
where we have applied Table~3.15 of Ref.~\cite{Schael:2013ita} taking only one of the Yukawa couplings to be non-zero.
In a similar way, one obtains the following constraints for the $\bar
e^c_R k^{++}e_R$ interaction: 
\begin{align}
& |h'_{ee}|\lesssim \frac{\sqrt{4\pi} m_{k^{\pm\pm}}}{8.6\ {\rm TeV}},\quad
 |h'_{e\mu}|\lesssim \frac1{\sqrt2}\frac{\sqrt{4\pi} m_{k^{\pm\pm}}}{11.6\ {\rm TeV}},\quad
 |h'_{e\tau}|\lesssim \frac1{\sqrt2}\frac{\sqrt{4\pi} m_{k^{\pm\pm}}}{8.7\ {\rm TeV}}.\end{align}

\subsection{The constraints from the ILC}
Here we discuss the constraints on the effective interactions
Eq.~(\ref{eq:effective1}) and (\ref{eq:effective2}) at the ILC. 
At the ILC, doubly charged scalar effects are studied in
\begin{align}
 & e^-(k_1,\sigma_1) + e^+(k_2,\sigma_2) \to e^-   (k_3,\sigma_3) +
 e^+(k_4,\sigma_4) , \\ 
 & e^-(k_1,\sigma_1) + e^+(k_2,\sigma_2) \to \mu^- (k_3,\sigma_3) +
 \mu^+(k_4,\sigma_4) , \\ 
 & e^-(k_1,\sigma_1) + e^+(k_2,\sigma_2) \to \tau^-(k_3,\sigma_3) +
 \tau^+(k_4,\sigma_4), 
\end{align}
where we assign momenta $k_i$ and helicities of initial~(final)-state
leptons $\sigma_i = \pm1$.
In each process, an additional diagram via the effective 4-point vertex
is added to the SM diagrams.

Applying Eq.~(\ref{eq:effective1}) and (\ref{eq:effective2}), helicity
amplitudes ${\cal M}_{\{\sigma_i\}}={\cal
M}(\sigma_1\sigma_2\sigma_3\sigma_4)$ for the first process are given 
as 
\begin{align}
  & {\cal M}(+-+-) = -e^2\left(1+\cos\theta\right)
 \left[ 1 + \frac{s}{t} + c_R^2\left(\frac{s}{s_Z}+\frac{s}{t_Z}\right)
 + \frac{2 s}{\alpha (\Lambda_R^e)^2}\right], \\ 
 & {\cal M}(-+-+) = -e^2\left(1+\cos\theta\right)
 \left[ 1 + \frac{s}{t} + c_L^2\left(\frac{s}{s_Z}+\frac{s}{t_Z}\right)
 + \frac{2 s}{\alpha (\Lambda_L^e)^2}\right], \\ 
 & {\cal M}(+--+) = {\cal M}(-++-) =
 e^2\left(1-\cos\theta\right)\left[1+c_Rc_L\frac{s}{s_Z}\right], \\
 & {\cal M}(++++) = {\cal M}(----) =
 2e^2\frac{s}{t}\left[1+c_Rc_L\frac{t}{t_Z}\right],
\end{align}
where $s=(k_1+k_2)^2=(k_3+k_4)^2$,
$t=(k_1-k_3)^2=(k_2-k_4)^2=-s(1-\cos\theta)/2$, 
$s_Z=s-m_Z^2+im_Z\Gamma_Z$, $t_Z=t-m_Z^2+im_Z\Gamma_Z$, 
and $\cos\theta$ is the scattering polar angle. 
Azimuthal angle dependence is neglected.
$e^2=4\pi\alpha$ with $\alpha$ being the QED coupling constant,
$c_R=\tan\theta_W$ and $c_L=-\cot2\theta_W$ where $\theta_W$ is the weak
mixing angle. 
The helicity amplitudes for the second and third processes are obtained
by removing terms with $1/t$ and $1/t_Z$ and replacing
$\Lambda_{L,R}^e \to\sqrt{2}\Lambda_{L,R}^{\mu,\tau}$. 

The differential cross-section for purely-polarized initial-state
($\sigma_{1,2}=\pm1$) is given as
\begin{align}
 \frac{d\sigma_{\sigma_1\sigma_2}}{d\cos\theta} = \frac{1}{32\pi s}
 \sum_{\sigma_3,\sigma_4} \left|{\cal M}_{\{\sigma_i\}}\right|^2.
\end{align}
The case for the partially-polarized initial-state with the degree of
polarization $P_{e^-}$ for the electron beam and $P_{e^+}$ for the
positron beam is given as
\begin{align}
  \frac{d\sigma(P_{e^-},P_{e^+})}{d\cos\theta} & =
 \frac{1+P_{e^-}}{2} \frac{1+P_{e^+}}{2}\frac{d\sigma_{++}}{d\cos\theta}
 + \frac{1+P_{e^-}}{2}
 \frac{1-P_{e^+}}{2}\frac{d\sigma_{+-}}{d\cos\theta}\nonumber \\
 & + \frac{1-P_{e^-}}{2}
 \frac{1+P_{e^+}}{2}\frac{d\sigma_{-+}}{d\cos\theta} 
 + \frac{1-P_{e^-}}{2}
 \frac{1-P_{e^+}}{2}\frac{d\sigma_{--}}{d\cos\theta}.
\end{align}
{ Polarized cross sections are useful to distinguish the left- and right-handed type interactions; e.g. a case of $(P_{e^-} > 0, P_{e^+} <0)$ is more sensitive to $\Lambda_R$ and less sensitive to $\Lambda_L$.}
For the realistic values at the ILC, we consider the following two
cases~\cite{Baer:2013cma}: 
\begin{align}
 & \frac{d\sigma_R}{d\cos\theta} = 
 \frac{d\sigma(0.8,-0.3)}{d\cos\theta}, \\
 & \frac{d\sigma_L}{d\cos\theta} = 
 \frac{d\sigma(-0.8,0.3)}{d\cos\theta},
\end{align}
defining polarized cross sections $\sigma_{R,L}$.
%

\begin{table}[t]
\begin{tabular}{c||c|c|c|c|c|c}
 \hline
 & \multicolumn{2}{c|}{$e^+e^-$} & \multicolumn{2}{c|}{$\mu^+\mu^-$} &
 \multicolumn{2}{c}{$\tau^+\tau^-$} \\
 \hline
 & $\Lambda_R^e$~[TeV] & $\Lambda_L^e$~[TeV] & $\Lambda_R^\mu$~[TeV] &
 $\Lambda_L^\mu$~[TeV] & $\Lambda_R^\tau$~[TeV] & $\Lambda_L^\tau$~[TeV] \\ 
 \hline
 \hline
 $\sigma_R$ & 42 & 12 & 63 & 17 & 60 & 16 \\
 \hline
 $\sigma_L$ & 11 & 39 & 14 & 60 & 13 & 56 \\
 \hline
\end{tabular}
 \caption{Upper limit on $\Lambda_{R/L}$ for $\sqrt{s}=500$~GeV with
 $L=1000$~fb$^{-1}$. \label{tab:limit1}} 
\end{table}
\begin{table}[t]
\begin{tabular}{c||c|c|c|c|c|c}
 \hline
 & \multicolumn{2}{c|}{$e^+e^-$} & \multicolumn{2}{c|}{$\mu^+\mu^-$} &
 \multicolumn{2}{c}{$\tau^+\tau^-$} \\
 \hline
 & $\Lambda_R^e$~[TeV] & $\Lambda_L^e$~[TeV] & $\Lambda_R^\mu$~[TeV] &
 $\Lambda_L^\mu$~[TeV] & $\Lambda_R^\tau$~[TeV] & $\Lambda_L^\tau$~[TeV] \\ 
 \hline
 \hline
 $\sigma_R$ & 21 & 6 & 30 & 8 & 28 & 8 \\
 \hline
 $\sigma_L$ & 6 & 21 & 7 & 28 & 6 & 26 \\
 \hline
\end{tabular}
 \caption{Upper limit on $\Lambda_{R/L}$ for $\sqrt{s}=250$~GeV with
 $L=250$~fb$^{-1}$. \label{tab:limit2}} 
\end{table}

\begin{figure}[t]
\begin{center}
\includegraphics[width=70mm]{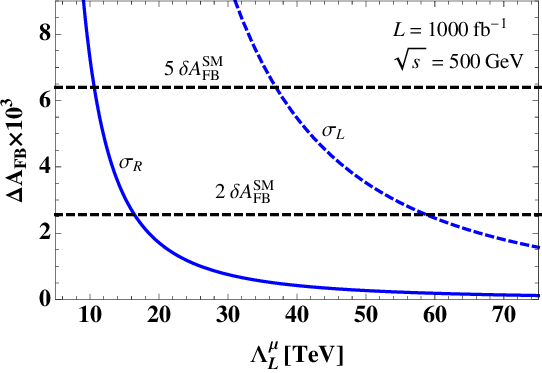} 
\includegraphics[width=70mm]{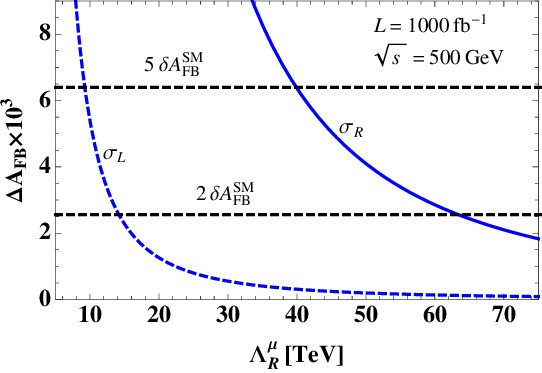} 
 \caption{Left (Right): $\Delta A_{FB}$ for $e^+ e^- \to \mu^+ \mu^-$ process as a function of $\Lambda_{L(R)}^\mu$ in which we take the other $\Lambda_{L,R}^{\ell}$ to be infinity by assuming corresponding Yukawa couplings to be zero, and $\sqrt{s} = 500$ GeV and $L = 1000$ fb$^{-1}$ are applied. }
 \label{fig:DelAFB}
\end{center}\end{figure}

To study the sensitivity to the doubly charged scalar effects, we consider
the measurement of a forward-backward asymmetry at the ILC, which is
defined as 
\begin{align}
 A_{FB} = \frac{N_F-N_B}{N_F+N_B}, 
\end{align}
where
\begin{align}
 & N_F = \epsilon \cdot L \cdot \int_0^{c_{\rm max}}
 d\cos\theta\frac{d\sigma}{d\cos\theta}, \\
 & N_B = \epsilon \cdot L \cdot \int_{-c_{\rm max}}^0
 d\cos\theta\frac{d\sigma}{d\cos\theta},
\end{align}
$L$ is an integrated luminosity, $\epsilon$ is an efficiency of
observing the events, and a kinematical cut $c_{\rm max}$ is chosen to
maximize the sensitivity.
For simplicity, we assume $\epsilon=100\%$ for $e^+e^-\to e^+e^-$
and $e^+e^-\to\mu^+\mu^-$, while $\epsilon=80\%$ for
$e^+e^-\to\tau^+\tau^-$~\cite{Tran:2015nxa}.
We take $c_{\rm max}=0.5$ for $e^+e^-\to e^+e^-$, and $c_{\rm max}=0.95$
for $e^+e^-\to \mu^+\mu^-$ and $\tau^+\tau^-$.

\begin{table}[t]
\begin{tabular}{c||c|c|c|c|c|c}
 \hline
 & \multicolumn{2}{c|}{$e^+e^-$} & \multicolumn{2}{c|}{$\mu^+\mu^-$} &
 \multicolumn{2}{c}{$\tau^+\tau^-$} \\
 \hline
 & $\Lambda_R^e$~[TeV] & $\Lambda_L^e$~[TeV] & $\Lambda^\mu_R$~[TeV] &
 $\Lambda^\mu_L$~[TeV] & $\Lambda^\tau_R$~[TeV] & $\Lambda^\tau_L$~[TeV] \\ 
 \hline
 \hline
 $\sigma_R$ & 35 & 10 & 53 & 14 & 50 & 13 \\
 \hline
 $\sigma_L$ & 9 & 33 & 12 & 50 & 11 & 47 \\
 \hline
\end{tabular}
 \caption{Upper limit on $\Lambda_{R/L}$ for $\sqrt{s}=500$~GeV with
 $L=500$~fb$^{-1}$.\label{tab:limit3}} 
\end{table}
\begin{table}[t]
\begin{tabular}{c||c|c|c|c|c|c}
 \hline
 & \multicolumn{2}{c|}{$e^+e^-$} & \multicolumn{2}{c|}{$\mu^+\mu^-$} &
 \multicolumn{2}{c}{$\tau^+\tau^-$} \\
 \hline
 & $\Lambda_R^e$~[TeV] & $\Lambda_L^e$~[TeV] & $\Lambda_R^\mu$~[TeV] &
 $\Lambda_L^\mu$~[TeV] & $\Lambda_R^\tau$~[TeV] & $\Lambda_L^\tau$~[TeV] \\ 
 \hline
 \hline
 $\sigma_R$ & 59 & 18 & 90 & 24 & 85 & 23 \\
 \hline
 $\sigma_L$ & 16 & 55 & 21 & 85 & 20 & 80 \\
 \hline
\end{tabular}
 \caption{Upper limit on $\Lambda_{R/L}$ for $\sqrt{s}=1$~TeV with
 $L=1000$~fb$^{-1}$.} 
 \label{tab:limit4}
\end{table}

{ A statistical error of the asymmetry for the SM is given by
\begin{align}
 \delta A_{FB}^{\rm SM} = \sqrt{\frac{1-(A^{\rm SM}_{FB})^2}{N_F^{\rm SM}+N_B^{\rm SM}}}.
\end{align}
where the values in the RHS are obtained by taking $\Lambda_{L,R}^\ell \to \infty$.}
We estimate the sensitivity to the scale $\Lambda_{R/L}^\ell$ ($2\sigma$
confidence level) by requiring
\begin{align}
{ \Delta A_{FB} \equiv } \left|A_{FB}(\Lambda)-A_{FB}^{\rm SM}\right|\ge 2 \delta
 A_{FB}^{\rm SM} .
 \label{eq:limitAFB}
\end{align}
For $\sqrt{s}=500$~GeV with an integrated luminosity $L=1000$~fb$^{-1}$,
we get the upper limit of $\Lambda^\ell_{R/L}$ which could be probed by using
$\sigma_R$ and $\sigma_L$ as summarized in Table~\ref{tab:limit1}; here we take one of $\Lambda^\ell_{L,R}$ is finite value and others are set to be infinity assuming corresponding Yukawa coupling is zero.
We thus find that the sensitivity to the scales of two types of
the effective interaction is significantly different for each polarized cross section. 
Therefore the type of the interaction can be distinguished by comparing the results for different beam polarizations.
{ To explicitly see the discrimination, we show $\Delta A_{FB}$ for $e^+ e^- \to \mu^+ \mu^-$ process as a function of $\Lambda^\mu_{L(R)}$ in left(right) plot of Fig.~\ref{fig:DelAFB} which is compared with $2 \sigma$ and $5 \sigma$ statistical errors for the SM.  We clearly see that the left(right)-handed type interaction leads $\sigma_{L(R)} \gg \sigma_{R(L)}$. In addition, we can discover the new physics effect with $5 \sigma$ significance when a value of $\Lambda_{L(R)}^\ell$ is not too large, and can distinguish a type of interaction by comparing $\sigma_L$ and $\sigma_R$. }

The scales we estimated for the case of the $\mu^+\mu^-$ final-state are
compared with those given in Ref.~\cite{ilc} where simulation studies of
discovering new physics effects by fitting the scattering angular
distribution are performed.
We find that their method improves the achievable scale by about
$20\%$-$30\%$ from our study using $A_{FB}$.
Therefore, we may expect such factors of improvement for the cases of
$e^+e^-$ and $\tau^+\tau^-$ final-states too.
Detailed study should be performed by including effects of beam-energy
spectrum, electroweak higher-order corrections, and detector
resolutions. 

{In addition, we present the results for different $\sqrt{s}$ and $L$ in
Tables~\ref{tab:limit2}, \ref{tab:limit3} and \ref{tab:limit4}. 
The upper limit of $\Lambda^\mu_{L(R)}$ can reach 85(90)~TeV for
$\sqrt{s}=1$~TeV with $L=1000$~fb$^{-1}$. 
Thus we obtain constraints on the couplings by $|h_{e \ell}|(|h'_{e
\ell}|) \leq \sqrt{2 \pi (1+\delta_{e \ell})} \times [ m_{\delta^{\pm
\pm} (k^{\pm \pm})}/ \Lambda^{\rm limit}_{L(R)} ]$. } 

\begin{figure}[t]
\begin{center}
\includegraphics[width=70mm]{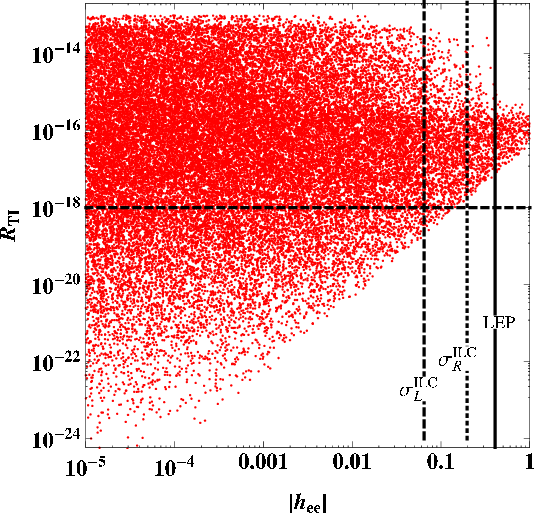} 
\includegraphics[width=70mm]{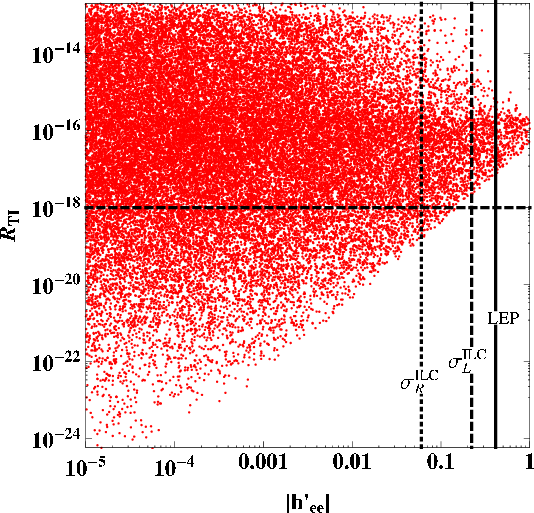} 
\includegraphics[width=70mm]{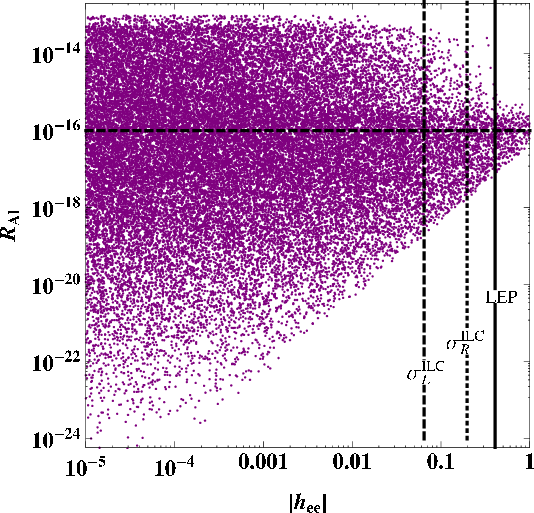} 
\includegraphics[width=70mm]{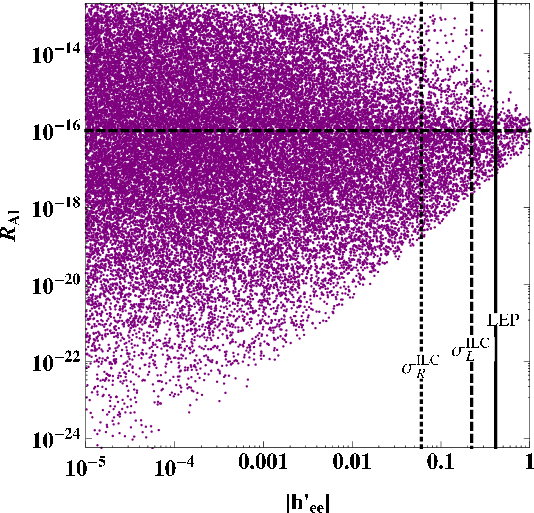} 
 \caption{Allowed regions of $|h_{ee}|$ versus $R_{\rm Ti(Al)}$ in the
 Higgs triplet interaction [upper~(lower) left panel], and $|h'_{ee}|$
 versus $R_{\rm Ti(Al)}$ in the Zee-Babu type interaction [upper~(lower)
 right panel] for fixed $m_{\delta^{\pm \pm}} = m_{\delta^\pm} =
 m_{k^{\pm \pm}} = 1$~TeV.
 Generated parameter points satisfying constraints from LFV processes
 are denoted by red and purple points for $R_{\rm Ti}$ and $R_{\rm Al}$
 respectively, vertical solid lines indicate LEP constraint, and
 dashed~(dotted) vertical lines represent the ILC bounds of the
 polarized cross sections $\sigma_{L(R)}$ for $\sqrt s= 1$~TeV and
 $L=1$~ab$^{-1}$ as shown in Table~\ref{tab:limit4}. 
 The horizontal lines represent the future upper bounds of $\mu-e$
 conversion.}
 \label{fig:hee-mue}
\end{center}\end{figure}
\begin{figure}[t]
\begin{center}
\includegraphics[width=70mm]{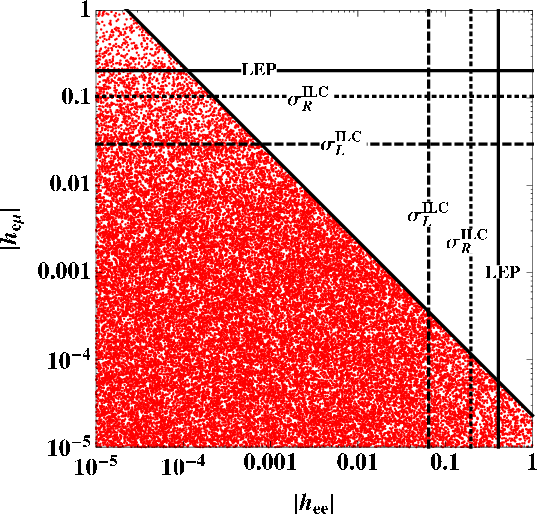} 
\includegraphics[width=70mm]{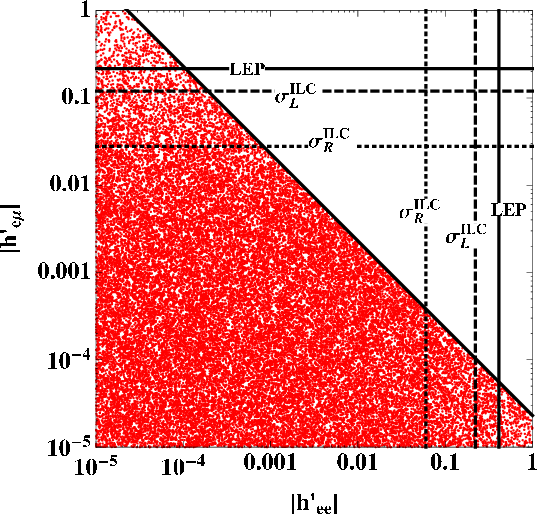} 
\includegraphics[width=70mm]{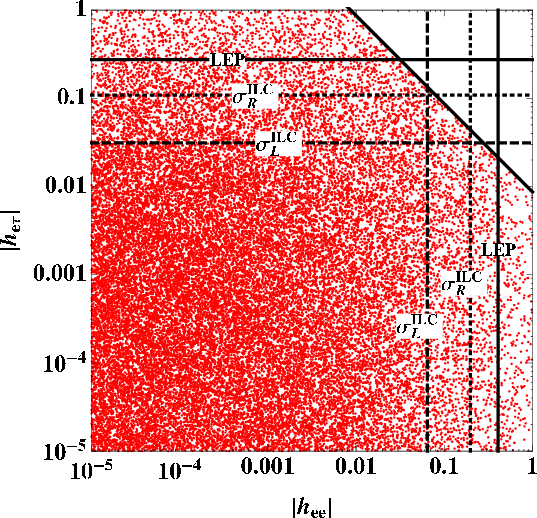} 
\includegraphics[width=70mm]{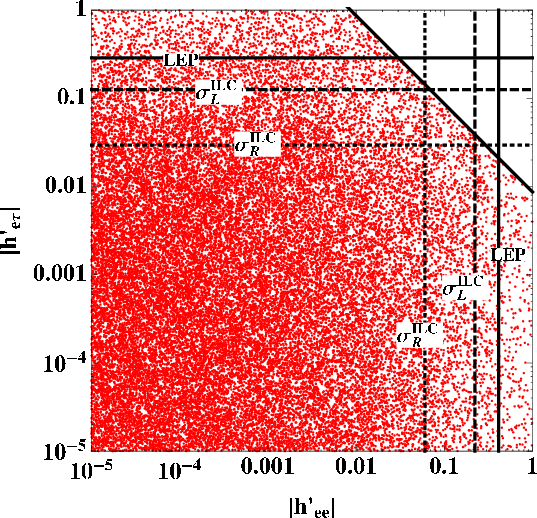} 
\caption{Allowed regions of $|h_{ee}|(|h'_{ee}|)$ versus $|h_{e
 \mu}|(|h'_{e \mu}|)$ [upper left~(right) panel]
 and $|h_{ee}|(|h'_{ee}|)$ versus $|h_{e\tau}|(|h'_{e \tau}|)$ [lower
 left~(right) panel] 
 for $m_{\delta^{\pm \pm}} = m_{\delta^\pm} = m_{k^{\pm \pm}} = 1$~TeV.
 Generated parameter points satisfying the LFV constraints are denoted
 by red points, the black solid lines represent the LEP bounds, 
 and the black dashed~(dotted) lines represent the ILC bounds of
 the polarized cross sections $\sigma_{L(R)}$ for $\sqrt s=1$~TeV and
 $L=1$~ab$^{-1}$.
Upper-right regions in the upper and lower figures are constrained by 
 $\mu\to 3e$ and $\tau\to 3e$, respectively.
 } \label{fig:hee-hij}
\end{center}\end{figure}

\section{Summary of the constraints with numerical analysis}

In this section, we summarize the constraints on the couplings
$h_{ij}$ and $h'_{ij}$ by combining bounds from the LFV processes and
the $e^+e^-\to\ell^+\ell^-$ processes.
{ We consider $10^5$ sets of Yukawa couplings $\{
|h_{ij}| \}( \{ |h'_{ij}| \})$ in the range of $\{10^{-5}, 1\}$ by
generating random numbers of $x_{ij}$ in $h_{ij}(h'_{ij})=10^{x_{ij}}$
in the range of $x_{ij} \in \{-5,0 \}$.
Then, we impose the constraints of LFV processes.
In the following, we consider fixed $m_{\delta^{\pm \pm}} =
m_{\delta^\pm} = m_{k^{\pm \pm}} = 1$~TeV, for simplicity. }

The upper~(lower)-left plot in Fig.~\ref{fig:hee-mue} shows the
allowed region in the $h_{ee}$-$R_{\rm Ti}(R_{\rm Al})$ plane while the
upper~(lower)-right plot shows that in the $h'_{ee}$-$R_{\rm Ti}(R_{\rm
Al})$ plane.
{ In these plots, each red~(purple) point indicates one
set of generated Yukawa couplings which satisfies the constraints of LFV
processes.}
{We note that; (I) absence of points in lower-right region in each plot
 is due to the lower bound in generating Yukawa couplings
 $|h_{ij}|(|h'_{ij}|) > 10^{-5}$, therefore artificial since the
 predictions on $R$ can be arbitrary small by setting small
 $h_{ij}(h'_{ij})$ except $h_{ee}(h'_{ee})$. 
 (II) constraints of LFV processes are slightly stronger on $|h_{ij}|$
 than $|h'_{ij}|$ because of contribution from the singly charged
 scalar. }
 We also indicate the LEP bound $|h_{ee}|(|h'_{ee}|) \lesssim 0.4$ and
 the ILC bound $|h_{ee}|(|h'_{ee}|) \lesssim 0.05 \sim 0.2$ depending
 on polarized cross sections for the case of $\sqrt{s} = 1$~TeV and
 $L=1000$~fb$^{-1}$; 
 the constraint is obtained by imposing Eq.~(\ref{eq:limitAFB}) where we apply Eq.~(\ref{eq:Lambda}) to extract bounds for Yukawa couplings. 
 In contrast to the constrains from LFV, we can obtain constraints on a single Yukawa coupling from the collider experiments.
 Note also that the constraints from $\sigma_L(\sigma_R)$ is stronger(weaker) than the other for left(right)-handed type interaction as we discussed in previous section. 
In addition, the future upper bound of the $\mu-e$ conversion rate is
shown by horizontal dashed line. 
 We find that within the parameter sets we
generated~($|h_{ij}|(|h'_{ij}|) \ge10^{-5})$ most of the regions where a signal
can be seen at the ILC can be checked by the $\mu-e$ conversion
experiment.
The ILC signal in the absence of the $\mu-e$ conversion signal may
indicate a hierarchical structure of the Yukawa matrix.

The upper~(lower)-left plot in Fig.~\ref{fig:hee-hij} shows the
allowed region in $|h_{ee}|$-$|h_{e\mu(e \tau)}|$ plane while the
upper~(lower)-right plot shows that in $|h'_{ee}|$-$|h'_{e\mu(e
\tau)}|$ plane. 
{In these plots, each red point corresponds to one of the generated
Yukawa coupling sets which satisfy the constraints from LFV processes. 
A lower density of points indicates that a larger fraction of
generated parameter sets are excluded in that region by the constraints
of the LFV processes other than $\mu(\tau) \to 3 e$.
The upper-right regions in the upper~(lower) plots are explicitly
excluded by the LFV processes $\mu(\tau) \to 3e$.} 
The LEP bounds are shown by black lines indicating $\{h_{ee}, h_{e \mu},
h_{e \tau} \} \lesssim \{ 0.40, 0.20, 0.26\}$  and $\{h'_{ee}, h'_{e
\mu}, h'_{e \tau} \} \lesssim \{ 0.40, 0.22, 0.28\}$ while the ILC
bounds by polarized cross section $\sigma_{L(R)}$ are shown by
dashed~(dotted) lines indicating $\{h_{ee}, h_{e \mu}, h_{e \tau} \}
\lesssim \{ 0.064(0.20), 0.030(0.10), 0.031(0.11) \}$ and $\{h'_{ee},
h'_{e \mu}, h'_{e \tau} \} \lesssim \{ 0.22(0.060), 0.12(0.028),
0.13(0.029) \}$ for the case of $\sqrt{s} = 1$~TeV and $L=1000$~fb$^{-1}$.
These bounds are simply scaled as $m_{\delta^{\pm \pm}, k^{\pm
\pm}}/{\rm TeV}$ for the different mass value of the doubly charged
scalar bosons. 
%

Here we comment on the Yukawa couplings $h_{\mu\mu} (h'_{\mu\mu})$, $h_{\tau\tau} (h'_{\tau\tau})$ and $h_{\mu\tau} (h'_{\mu \tau})$. 
{These couplings are not shown in Figs.~\ref{fig:hee-mue} and \ref{fig:hee-hij} explicitly, since they are not directly constrained by lepton collider experiments in our analysis.
However they are constrained from the LFV processes as Tables~\ref{tab:LFVconst1}, \ref{tab:LFVconst2} and \ref{tab:LFVconst3} combining with other couplings.
Then we have implicitly included LFV constraints when we run these parameters in generating Figs.~\ref{fig:hee-mue} and \ref{fig:hee-hij}. }
We also note that flavor-violating scattering processes, $e^+ e^- \to
\ell^+ \ell'^-$, can also be investigated at the ILC~\cite{Cho:2016zqo}
which will test combinations of the Yukawa couplings such as $h_{ee}
h_{e \mu (\tau)}$.

{Note that including fit to neutrino oscillation data LFV and LHC physics are investigated for effective theory obtained by integrating out {the SM charged-leptons, an $SU(2)$ singlet singly-charged scalar, and} doubly-charged scalar in Ref.~\cite{King:2014uha}. Then the relative size of Yukawa couplings are constrained from neutrino oscillation data and LHC can search for doubly charged Higgs where the signal is determined by branching ratio for decay process $\delta^{\pm \pm} \to \ell^\pm_i \ell^\pm_j$. If we find the signal of doubly charged Higgs at the LHC, the relative magnitude of Yukawa couplings can be investigated via branching ratio. One advantage of our analysis for lepton collider is that we can constrain absolute magnitude of some Yukawa couplings as we discussed above. }

{In our analysis we have not included fitting with neutrino oscillation data to investigate direct constraints on the Yukawa couplings associated with doubly charged scalar in general. 
If we take into account neutrino oscillation data, relative size of Yukawa couplings are constrained; absolute sizes of Yukawa couplings are controlled by VEV of triplet in type-II seesaw case and by mass scale inside loop diagram in Zee-Babu type models. For example, in type-II seesaw case we obtain the relation $h_{11} >(<) \, h_{22,33}$ for normal(inverted) ordering neutrino mass~\cite{Perez:2008ha}. Also some constraints on relative sizes of Yukawa couplings can be obtained in Zee-Babu model from fitting to neutrino oscillation data~\cite{Herrero-Garcia:2014hfa}, which are weaker than the case of type-II seesaw since we have freedom to tune Yukawa couplings associated with singly charged scalar field.
Therefore combining neutrino oscillation data and tests for effective operators at the ILC, we can further explore the parameter region for Yukawa couplings by specifying neutrino mass generation mechanism. }


\section{Conclusions}
We have explored discrimination of two types of leptonic Yukawa
interactions associated with a Higgs triplet or $SU(2)$-singlet doubly
charged scalar boson where the former one appears in realization of
type-II seesaw mechanism for neutrino mass generation and the latter 
one appears in Zee-Babu type models of radiative neutrino mass generation.

First we have reviewed the constraints from lepton flavor violating~(LFV)
processes for each type of Yukawa interactions where $\ell_i \to \ell_j
\ell_k \ell_l$, $\ell \to \ell' \gamma$ and $\mu-e$ conversion
processes are discussed. 
Next we have shown that these interactions can be distinguished at the
ILC by measuring the difference of the scattering angular distribution
in the $e^+e^-\to\ell^+\ell^-$ processes with polarized
electron and positron beams due to the chirality difference of the
interactions.
The forward-backward asymmetry in the scattering angular distribution
has been investigated to obtain the upper bound of these couplings.
Finally we have shown prospects of bounds on the model parameter space by
combining the constraints of future LFV processes and the
$e^+e^-\to\ell^+\ell^-$ processes at the ILC.

\section*{Acknowledgments}
\vspace{0.5cm}
H.~O.\ is sincerely grateful for all the KIAS members. 

\end{document}